\documentclass[twocolumn,preprintnumbers,amsmath,nofootinbib,amssymb]{revtex4}
\usepackage{amssymb}
\usepackage{latexsym}
\usepackage{color}
\usepackage{graphicx}
\usepackage{enumitem}  
\usepackage{hyperref}
\usepackage{color}

\def\nn{\nonumber\\}
\newcommand{\f}[2]{\frac{#1}{#2}}
\def\be{\begin{equation}}
\def\ee{\end{equation}}
\def\bea{\begin{eqnarray}}
\def\eea{\end{eqnarray}}
\begin{document}
\title{Non-Singular Collapse Scenario From Matter-Curvature Coupling}
\author{A. H. Ziaie$^1$\footnote{ah.ziaie@maragheh.ac.ir}, H. Moradpour$^1$\footnote{hn.moradpour@maragheh.ac.ir}, M. Mohammadi Sabet$^2$\footnote{m.mohamadisabet@ilam.ac.ir},}
\address{$^1$ Research~Institute~for~Astronomy~and~Astrophysics~of~ Maragha~(RIAAM), University of Maragheh, P.~O.~Box~55136-553,~Maragheh, Iran\\$^2$ Basic Science Faculty, Physics department, Ilam University, P. O. Box, 69315-516, Ilam, Iran}
\begin{abstract}
In the present work we study spherically symmetric gravitational collapse of a homogeneous perfect fluid in the context of Generalized Rastall Theory (GRT). In this modified version of the original {Rastall Gravity (RG)}, the coupling parameter which is a representative of matter-curvature interaction is no longer a constant parameter. Such a dynamic coupling may play the role of dark energy which is responsible for the present accelerating expansion of the Universe. Assuming then a linear equation of state (EoS) for the fluid profiles, we seek for physically reasonable collapse scenarios in which the spacetime singularity that occurs in general relativity (GR) is replaced by a non-singular bounce. We therefore find that depending on model parameters, the collapse process which starts from regular initial data, will halt at a minimum value for the scale function and then turns into an expansion at a finite time. We further find that there exists a minimum value for the initial radius of collapsing object so that for radii smaller than this minimum radius, formation of apparent horizon can be avoided and hence the bounce can be visible to the observers within the Universe. We also compare our results to quantum corrected collapse scenarios and find that the mutual interaction between matter and geometry can play the role of quantum corrections to energy density.
\end{abstract}

\maketitle
\section{Introduction}
The final fate of gravitational collapse of dense body under its own gravity is an important problem of relativistic astrophysics and gravitational physics. In the framework of GR, the Hawking and Penrose singularity theorems predict that under physically reasonable conditions, the collapsing object undergoes a continuous contraction, reaching then higher and higher densities and curvatures and finally the collapse process ends in the formation of a curvature singularity. {It is generally believed that in the vicinity of such an extreme spacetime event} densities, spacetime curvatures and all other physical quantities diverge and classical framework of the theory breaks down~\cite{HAWPENST,HAWPENST1}, {see also~\cite{Clarke-Tipler,Senovilla1997} for recent review}. In other words, the behaviour of such extreme regions may not be governed by the classical GR theory itself, and a quantum theory of gravity would be the most likely description of the phenomena created by such spacetime events~\cite{QG}.
\par
The first attempt towards understanding gravitational collapse processes in relativistic astrophysics was carried out by Datt~\cite{DATT1938} in 1938 and later by Oppenheimer and Snyder~\cite{OS1939} in 1939, where general relativity was utilized to study the dynamical collapse of a homogeneous spherical dust cloud under its own weight. According to this model, which is now known as OSD collapse, a homogeneous dust collapse leads to the formation of a black hole within the spacetime. In this case, the spacetime singularity that forms as the collapse end product is hidden behind an event horizon and thus is causally disconnected from other regions of spacetime, as hypothesized by the cosmic censorship conjecture (CCC)~\cite{CCC} (See also~\cite{CCC1} for reviews on the conjecture). However, it was realized that GR field equations also admit naked singularities, i.e., spacetime events where super-dense regions of extreme gravity can be detected by faraway observers in the Universe, in violation to CCC~\cite{COLLQUANNSIN,COLLQUANNSIN1}. The existence of such objects in GR has been predicted under a variety of physical circumstances with matter content of various types~\cite{NSSF} and in the context of modified gravity theories~\cite{NSMODG} (see also~\cite{NSREV} for some recent reviews).
\par
{Despite} its successes~\cite{GRtest}, GR generically suffers from the issues caused by the occurrence of spacetime singularities, e.g., problems such as future predictability, path incompleteness and etc.~\cite{GRSIN}. This is due to the fact that when we model {a} physical system in the framework of classical GR, solutions of the field equations can predict the evolution of the system. Therefore, if it happens that at some spacetime event physically relevant quantities grow unboundedly and diverge, we {may interpret such an ill-behavior as occurrence of a singularity}. Practically, as the classical framework of GR loses its application when basic quantities become infinite, {formation of a singularity may be regarded as} a sign that the theory has been employed beyond its domain of validity~\cite{GRSIN1}. Thus, in order to remedy the singularity problem, it is reasonable to search for alternative theories of gravity whose geometrical attributes are not present in GR. Many efforts have been made in this direction and the results of research show that the presence of additional/correction terms within the classical framework of modified gravities can provide a setting to avoid singularities within the framework of GR. For example, in $f(R)$ gravity, authors of~\cite{fRnonsin} have studied a curvature singularity occurring in the collapse process and found that addition of term $\propto R^\alpha$ ($1<\alpha<2$) could remove the curvature singularity. Moreover, the possibility of singularity removal in generalized Brans-Dicke theory with a running coupling function has been discussed in~\cite{NONSINBD}. The effects of intrinsic angular momentum (spin of fermionic matter) on the collapse endstate have been studied in the context of Einstein-Cartan theory~\cite{ECnonsin}, where it is shown that the effects of a spin source can replace the spacetime singularity by a non-singular bounce. Also, in the framework of Einstein-Cartan gravity it is shown that a non-minimal coupling between gravity and fermions can alter the final fate of classical collapse process long before quantum gravity effects become dominant~\cite{4fermcoll}. However, it is widely believed that in very late stages of a collapse scenario where the Planck scale physics become important, quantum effects of gravity come into play to finally remove the classical singularity~\cite{COLLQUANNSIN}. In this regard, non-singular quantum collapse models have been vastly investigated and the results show that Planck scale effects can avoid the spacetime singularity, replacing it by a non-singular quantum bounce~\cite{ALTGRNONSIN,Tavakoli2014,Bambi2013}.
\par
Most of alternative theories of gravity respect the conservation of energy-momentum tensor (EMT) which is expressed by the zero divergence of this tensor. In other words, this conservation law conveys minimal coupling of matter source to the geometry. Nevertheless, one more possible way to generalize GR beyond its constraint on EMT conservation is to relax this condition, so that the ordinary energy-momentum conservation law ($\nabla_\mu T^\mu_{\,\,\,\, \nu}=0$) does not hold anymore. Indeed, the $\nabla_\mu T^\mu_{\,\,\,\, \nu}\neq0$ condition has been phenomenologically confirmed by particle production process in cosmology~\cite{ppp0,PPP}. The first attempt towards developing this idea in GR framework comes back to the Rastall's work~\cite{rastall}, which states that, conservation laws have been only tested on the Minkowski spacetime or quasistatic gravitational fields~\cite{rastall,od1,od2,cmc,cmc1,cmc2,genras}. Based on Rastall's argument, the covariant derivative of EMT is proportional to the gradient of the Ricci curvature scalar, i.e., $\nabla_\mu T^\mu_{\,\,\,\, \nu}=\lambda\nabla_\mu{\mathcal R}$ where $\lambda$ is a constant coupling parameter. From another sight, RG allows the geometry and energy-momentum sources be coupled to each other (with a constant coupling $\lambda$ parameter) in a non-minimal way~\cite{genras,CMCRAS}, a result which also emerges from curvature-matter theories of gravity~\cite{rastall,CMC,cmc1,cmc2}. Such a mutual interaction can give rise to many interesting consequences, namely, interesting and new unexpected aspects of the cosmological models in {RG} have been investigated in~\cite{IJMPCS}. Furthermore, it has been shown that Rastall's modification of GR is in good agreement with observations as well as theoretical expectations~\cite{genras,Effemtras}. More recently, a generalization of {RG} has been introduced where it is shown that a time-varying coupling parameter, i.e., $\lambda(t)$, can play the role of dark energy and thus, the current accelerating phase of Universe may have been originated from such a dynamic matter-geometry interaction~\cite{genras}. This model successfully describes the evolution of the Universe and cosmic acceleration. Finally, the authors of~\cite{MGRASQian} have considered a more general approach for extending the {RG}. They assumed a second rank tensor field which is proportional to spacetime metric and a function of Ricci curvature scalar and the trace of EMT. The non-zero covariant divergence of this tensor field is then considered as the non-conserved sector of the EMT and consequently acts as a dark energy source during the cosmic evolution. It is therefore found that during the dark energy dominated era the amount of violation of EMT is more considerable.
\par
Motivated by the above arguments, our aim in the present work is to examine the possibility of singularity avoidance in gravitational collapse of a homogeneous perfect fluid within the framework of GRT and seek for possible effects of a running mutual interaction between matter and geometry on the collapse dynamics and its final outcome. The paper is then organized as follows: In Sec.~\ref{OSD} we give a brief review on the Oppenheimer-Snyder collapse model. Sec.~\ref{NSGRT} deals with non-singular collapse setting in GRT and Sec.~\ref{QUGRT} is devoted to a comparison between the current model and quantum corrected collapse ones. Our conclusions are drawn in Sec.~\ref{ConC}. We set the units so that $c=8\pi G=1$.
\section{A Brief Review on OSD collapse model}\label{OSD}
In the framework of classical GR, the simplest model describing gravitational collapse of a homogeneous and marginally bound dust cloud is that of the OSD collapse model~\cite{DATT1938,OS1939}. The end product of such a process is the formation of a black hole since during the {collapse evolution}, a dynamical horizon forms to cover the spacetime singularity, hence, the singularity is hidden from all observers at infinity. The dynamic horizon coincides with event horizon when the collapse process settles down to a static configuration. The interior spacetime of such a collapse setting can be described by a spatially closed Friedman-Robertson-Walker metric given by 
\be\label{FLRW}
ds^2=-d\tau^2+\f{a^2(\tau)}{1-r^2}dr^2+R^2(\tau,r)d\Omega^2,
\ee
where $R(\tau,r)=ra(\tau)$ is the physical radius of the collapsing object, with $a(\tau)$ being the scale factor and $d\Omega^2$ is the standard line element on the unit 2-sphere. The EMT of a dust fluid is simply given by $T^{a}_{\,\,b}={\rm diag}(-\rho,0,0,0)$, with the help of which the Einstein field equations read
\bea
\f{3}{a^2}+3\f{\dot{a}^2}{a^2}=\rho,\label{EFEsdust0}\\
\f{1}{a^2}+\f{\dot{a}^2}{a^2}+2\f{\ddot{a}}{a}=0,\label{EFEsdust1}
\eea
Moreover, conservation of EMT ($\nabla_\alpha T^\alpha_\beta=0$) leads to $\rho=\rho_{\rm i}({a_{\rm i}}/{a})^{3}$ where $\rho_{\rm i}$ is the initial value of the {energy density} at initial epoch, i.e., when $a(\tau_{\rm i})=a_{\rm i}$. Substituting the expression for energy density into Eq. (\ref{EFEsdust0}) together with defining the conformal time $d\eta=d\tau/a$, we find the following solution for the scale factor
\be\label{solosd}
a(\eta)=\f{a_i}{2}(1+\cos(\eta)),~~~\tau(\eta)=\f{a_i}{2}(\eta+\sin(\eta)),
\ee
where $0\leq\eta\leq\pi$. The above solution describes the process of a homogeneous dust cloud collapse so that the scale factor begins its evolution from the finite value $a_{\rm i}$ at $(\tau_{\rm i},\eta_{\rm i})=(0,0)$ and becomes zero at $(\tau_{\rm s},\eta_{\rm s})=(\pi a_i/2,\pi)$. We observe that the scale factor vanishes at a finite amount of time, thus the collapse ends in a spacetime singularity. In order to check whether this singularity is visible or not we need to investigate the behavior of apparent horizon. To this aim we define the null coordinates~\cite{Hay1996}
\bea\label{nullcoor}
d\zeta^+&=&-\f{1}{\sqrt{2}}\left[d\tau-\f{a}{\sqrt{1-r^2}}dr\right],\nn d\zeta^-&=&-\f{1}{\sqrt{2}}\left[d\tau+\f{a}{\sqrt{1-r^2}}dr\right],	
\eea
with the help of which the line element (\ref{FLRW}) can be recast into the following double-null form as
\be\label{metdoublen}
ds^2=-2d\zeta^+d\zeta^-+R^2(\tau,r)d\Omega^2.
\ee
The radial null geodesics are given by the condition $ds^2=0$. Thus, there exists two kinds of null geodesics corresponding to $\zeta^+={\it constant}$ and $\zeta^-={\it constant}$. The expansion parameters along these null geodesics are {then} given by
\be\label{Thetapm}
\Theta_{\pm}=\f{2}{R(\tau,r)}\f{\partial}{\partial\zeta^{\pm}}R(\tau,r),
\ee
where 
\bea
\f{\partial}{\partial\zeta^{+}}&=&\f{1}{\sqrt{2}}\left[\partial_\tau+\f{\sqrt{1-r^2}}{a}\partial_r\right],\label{zetap}\\
\f{\partial}{\partial\zeta^{-}}&=&\f{1}{\sqrt{2}}\left[\partial_\tau-\f{\sqrt{1-r^2}}{a}\partial_r\right],\label{zetam}
\eea
denote coordinate derivatives along $\zeta^{\pm}$. The expansion parameters measure whether the light beams normal to a sphere are diverging $\Theta>0$ or converging $\Theta<0$, or equivalently, whether the area of the spheres is increasing or decreasing along the null directions. The spacetime is said to be trapped, untrapped and marginally trapped if~\cite{Hay1996}
\be\label{THtruntr}
\Theta_+\Theta_->0,~~~~\Theta_+\Theta_-<0,~~~~\Theta_+\Theta_-=0,
\ee
respectively, where the third case implies the outermost boundary of the trapped region, i.e., the apparent horizon. From expressions (\ref{Thetapm})-(\ref{zetam}) we get
\bea\label{THPM}
\Theta_+\Theta_-&=&\f{2}{R^2}\left[r^2(1+\dot{a}^2)-1\right]\nn
&=&\f{2}{R^2}\left[\f{\rho_{\rm i}a_{\rm i}^3R_{\rm i}^2}{a}-1\right],
\eea
where use has been of the coordinate freedom $R(\tau_{\rm i},r)=R_{\rm i}=r$ to rescale the radial coordinate $r$ as the initial radius of the collapsing body. The apparent horizon can then be located through the condition $\Theta_+\Theta_-=0$, which gives
\be\label{APHOR}
\eta_{\rm ah}=\cos^{-1}\left(2\rho_{\rm i}a_{\rm i}^2R_{\rm i}^2-1\right).
\ee
We therefore observe that since the initial radius is nonzero, $\eta_{\rm ah}<\eta_{\rm s}$ always, unless in the limit where $R_{\rm i}\rightarrow0$ the apparent horizon formation coincides with singularity time, i.e., $\eta_{\rm ah}\rightarrow\eta_{\rm s}=\pi$. However, a zero initial radius of the collapsing cloud is unphysical. We then conclude that the apparent horizon forms earlier than the singularity formation and thus a black hole will arise as the end-product of a homogeneous dust collapse, see also~\cite{OSDreview} for a review on OSD model. 
\section{Non-Singular Collapse Model In GRT}\label{NSGRT}
Our aim in the present section is to build and study non-singular homogeneous collapse models in the framework of GRT and seek for possible effects of a running coupling parameter on the collapse end product. In GRT, the original assumption of Rastall~\cite{rastall} is extended to include a varying coupling parameter. The non-conservation of EMT then reads
\be\label{CoVDiv}
\nabla_{a}T^{a}_{\,\,\,b}=\nabla_{b}({\lambda\mathcal R}),
\ee
where, in general, the $\lambda$ parameter depends on spacetime coordinates. With the help of Bianchi identity $\nabla^{\nu}G_{\mu\nu}=0$ we get the GRT field equations as~\cite{genras}
\be\label{RastallFES}
G_{ab} +\kappa\lambda g_{ab}{\mathcal R}=\kappa T_{ab},
\ee
where $\kappa$ is a constant. We note that though the above equation is similar to the field equations of {RG}, here $\lambda$ is not generally a constant parameter. The field equations for an isotropic source ($T^{a}_{\,\,b}={\rm diag}(-\rho,p,p,p)$) {are then given by}
\bea
\!\!\!\!\!\!\!\!\!\!\!\!\!\!\!\!\!\!&&6\kappa\lambda\f{\ddot{a}}{a}+3(2\kappa\lambda-1)\left(\f{\dot{a}}{a}\right)^2\!\!+\f{3}{a^2}(2\kappa\lambda-1)\!\!=\!\!-\kappa\rho,\label{RasFES}\\
\!\!\!\!\!\!\!\!\!\!\!\!\!\!\!\!\!\!&&2(3\kappa\lambda-1)\f{\ddot{a}}{a}+(6\kappa\lambda-1)\left(\f{\dot{a}}{a}\right)^2\!\!+\!\f{1}{a^2}(6\kappa\lambda-1)=\kappa p.\label{RasFES1}
\eea
For the continuity equation (\ref{CoVDiv}) we get
\be\label{conteq}
\f{d}{d\tau}\left(\rho+\lambda{\mathcal R}\right)+3H(\rho+p)=0,
\ee
where $H=\dot{a}/a$ is the collapse rate and $\dot{}\equiv d/d\tau$. We assume that the collapsing material obeys a linear equation of state given as $p=w\rho$. Hence, in order to better deal with the collapse solutions for such type of fluid we firstly change the derivatives from co-moving time to scale factor using the relation
\be\label{Ha}
\dot{H}=\f{dH}{d\tau}=aH(a)H^\prime(a),
\ee
where $\prime\equiv d/da$. Eqs. (\ref{RasFES}), (\ref{RasFES1}) and (\ref{conteq}) can then be rewritten as
\bea
3\left[1-4\kappa\lambda(a)\right]H^2(a)-6\kappa\lambda(a)aH(a)H^\prime(a)\nn+\f{3\left[1-2\kappa\lambda(a)\right]}{a^2}-\kappa\rho(a)=0,\label{RasFESa}\\
3\left[1-4\kappa\lambda(a)\right]H^2(a)+2\left[1-3\kappa\lambda(a)\right]aH(a)H^\prime(a)\nn+\f{\left[1-6\kappa\lambda(a)\right]}{a^2}+\kappa w\rho(a)=0,\label{RasFESa1}\\
a\rho^\prime(a)+6a\lambda^\prime(a)\left[2H^2(a)+aH(a)H^\prime(a)+\f{1}{a^2}\right]\nn+6a\lambda(a)\Big[5H(a)H^\prime(a)+a\left(H^\prime(a)\right)^2\nn+aH(a)H^{\prime\prime}(a)-\f{2}{a^3}\Big]+3(1+w)\rho(a)=0.\label{RasFESacons}
\eea
The above system of equations admits a general solution given as
\bea
\rho(a)\!\!&=&\!\!C_1{\exp}{\int f(a)da},\label{rhoasol}\\
\!\!H^2(a)\!\!&=&\!\!\f{\kappa\lambda(a)\left[3\kappa a^2(1+w)\rho(a)-12\right]-\kappa\rho(a)a^2+3}{12\kappa\lambda(a)a^2-3a^2},\nn\label{H2asol}
\eea
where
\be
f(a)=\left[\f{a\kappa(3w-1)\lambda^\prime(a)-3(1+w)(4\kappa\lambda(a)-1)^2}{a(4\kappa\lambda(a)-1)(3\kappa(1+w)\lambda(a)-1)} \right],
\ee
and $C_1$ is an integration constant. Next, we proceed to build and study collapse scenarios assuming $f(a)=\beta/a$ where $\beta$ is a constant. We therefore obtain the following expression for $\lambda(a)$ parameter as
\be\label{lambdasol}
\lambda(a)=\f{a^\beta(3w+\beta+3){\rm e}^{\beta C_1}-1}{3\kappa a^\beta(\beta+4)(1+w){\rm e}^{\beta C_1}-4\kappa}.
\ee
Substituting the above solution into Eq.(\ref{rhoasol}) the integration can be readily performed with the following solution for energy density
\bea
\rho(a)&=&\rho_{\rm i}\left(\f{a}{a_{\rm i}}\right)^\beta,\label{energa}
\eea
where $\rho_{\rm i}$ is the initial energy density of the collapsing body at initial time $\tau_{\rm i}$ where $a(\tau_{\rm i})=a_{\rm i}$. Using the above solution along with Eq.(\ref{lambdasol}), expression (\ref{H2asol}) for the collapse rate reads
\be\label{collrate}
H^2(a)=\f{\beta(a^2-a_{\rm b}^2)a_{\rm i}^\beta+a^2a_{\rm b}^2\rho_{\rm i}\kappa(1+w)(a_{\rm b}^\beta-a^\beta)}{\beta a_{\rm b}^2a_{\rm i}^\beta a^2}.
\ee
The integration constant in Eq. (\ref{lambdasol}) has been determined in such a way that the collapse process halts at the bounce time, i.e., $H(a_{\rm b})=0$ where, $a_{\rm b}=a(\tau_{\rm b})$ is the value of the scale factor at the bounce time $\tau_{\rm b}>\tau_{\rm i}$. Clearly we must have $a_{\rm b}<a_{\rm i}$. Substituting then for $C_1$ into Eq. (\ref{lambdasol}) we finally get
\bea\label{lambdasolfinal}
&&\lambda(a)=\nn&&\f{3\beta a_{\rm i}^\beta+3\kappa a_{\rm b}^{\beta+2}(1+w)\rho_{\rm i}-\kappa(\beta+3w+3)\rho_{\rm i}a_{\rm b}^2a^\beta}{12\kappa\left(\beta a_{\rm i}^\beta+\kappa\rho_{\rm i}(1+w)a_{\rm b}^{\beta+2}-\f{\kappa}{4}\rho_{\rm i}a_{\rm b}^2(\beta+4)(1+w)a^\beta\right)}.\nn
\eea
It is worth noting that the above solution is verified using the trace of field equation Eq.~(\ref{RastallFES}). A straightforward calculation then gives
\bea\label{lambtrace}
\lambda&=&\f{1}{4\kappa}\left(1+\f{\kappa(3w-1)\rho}{{\mathcal{R}}}\right),
\eea
whence using Eqs.~(\ref{Ha}) and (\ref{energa}) to substitute for energy density and Ricci scalar we obtain the same expression for $\lambda(a)$ parameter as given in Eq. (\ref{lambdasolfinal}). We further note that since $\rho$ is the representative of mass-energy distribution within the spacetime and ${\mathcal{R}}$ is that of spacetime curvature, the ratio $\rho/{\mathcal{R}}$ can be regarded as the mutual interaction between matter and curvature that the measure of which is described by $\lambda$ parameter.
\par
\begin{figure}
	\includegraphics[scale=0.27]{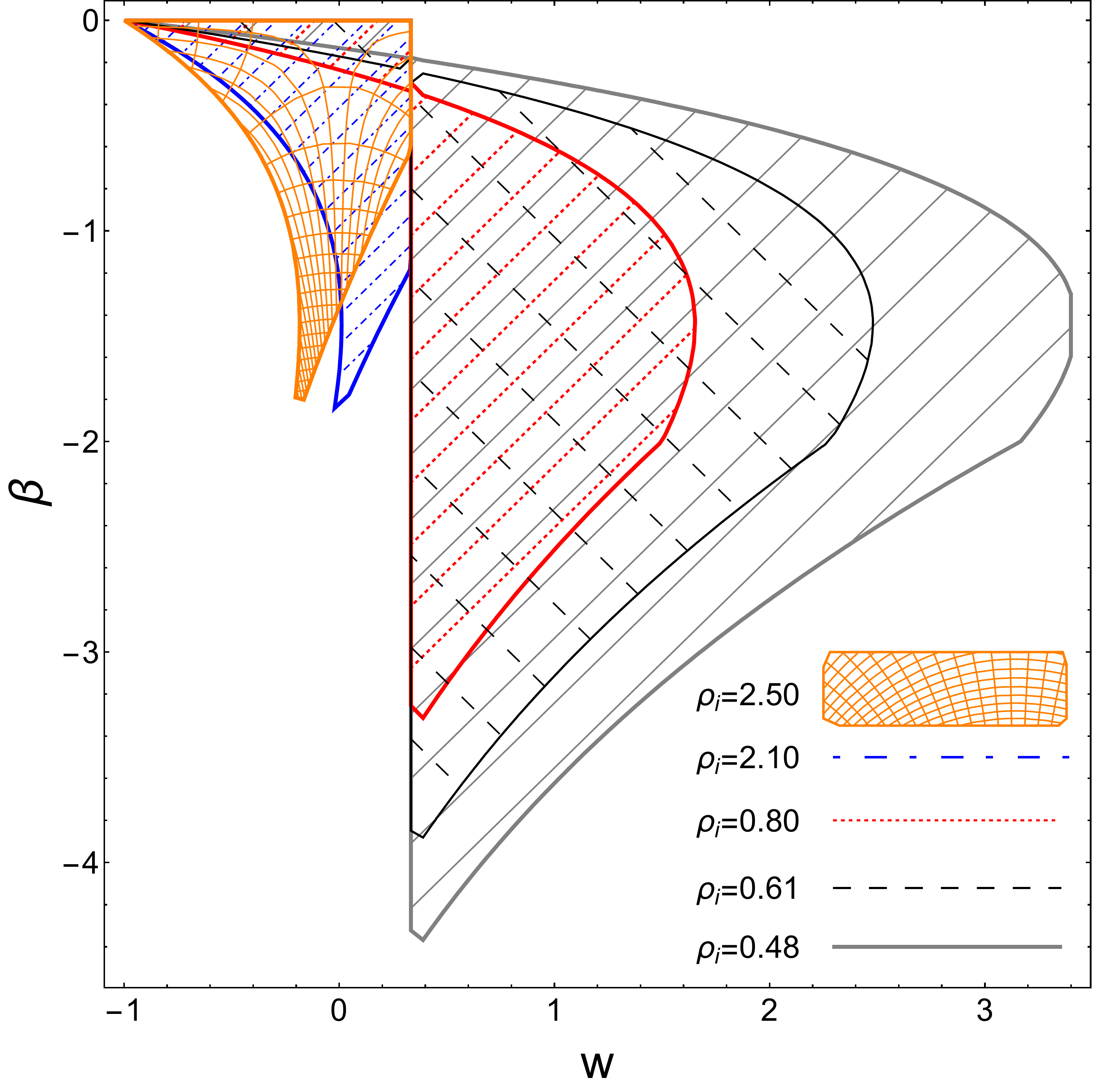}
	\caption{The allowed values for EoS and $\beta$ parameters for $a_{\rm i}=2a_{\rm b}=1$, $\kappa=1$, $R_{\rm i}=0.1$ and different values of initial energy density.}\label{FIG1}
\end{figure}
\begin{figure}
	\includegraphics[scale=0.3]{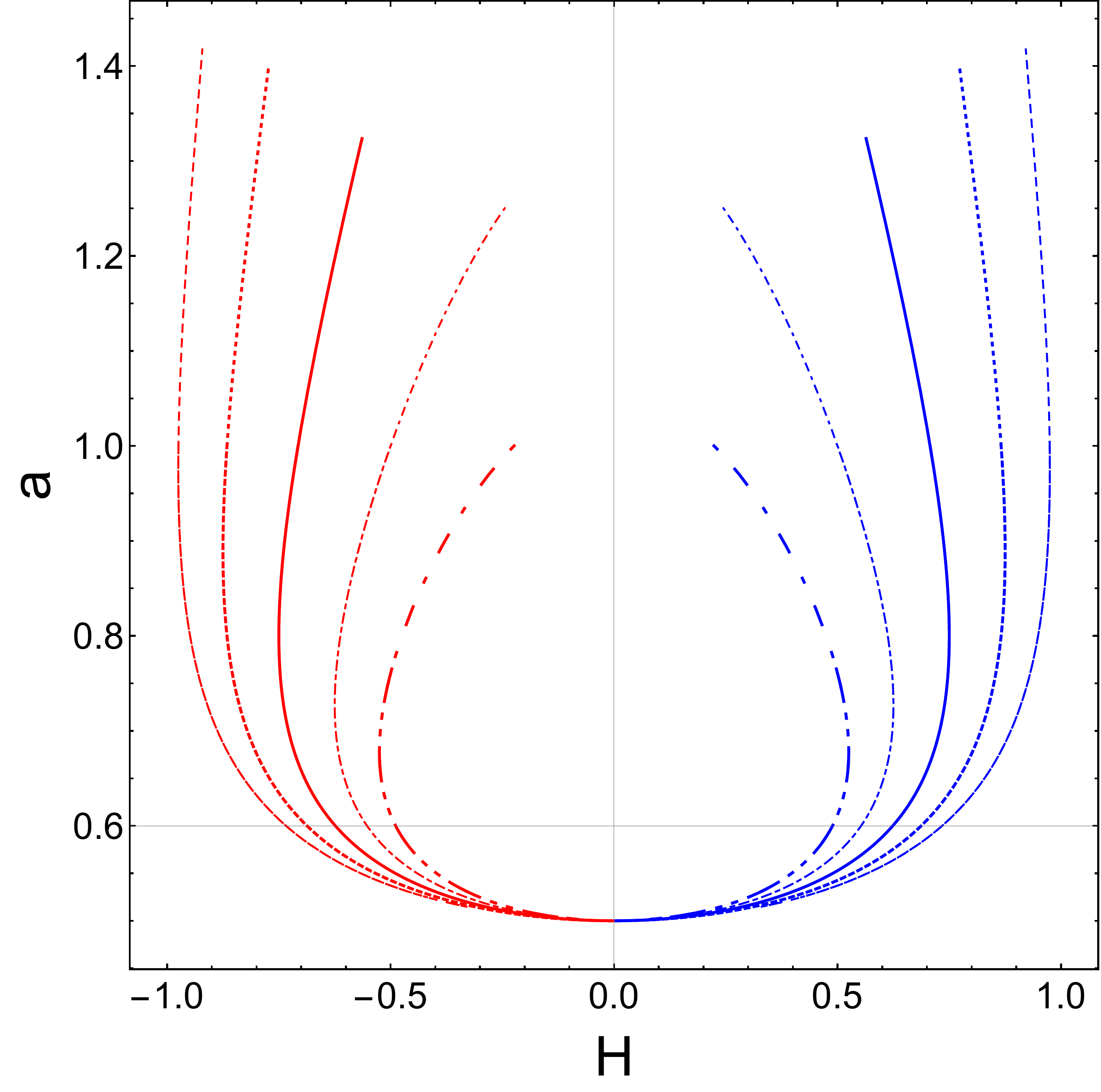}
	\caption{Evolution of collapse rate against scale factor for $\beta=-1$, $w=-0.18$ (dashed curves), $w=-0.10$ (dotted curves), $w=0$ (solid curves), $w=0.10$ (dotdashed curves) and $w=0.18$ (long-dotdashed curves) and for $a_{\rm i}=2a_{\rm b}=1$ and $\kappa=1$. The family of Red curves represent a contracting regime and those of Blue curves show an expanding one.}\label{FIG2}
\end{figure}

In order to have physically reasonable non-singular solutions we put the following conditions on our collapse setting:
\begin{enumerate}[label=(\roman*)]
	\item The collapse begins from regular initial data defined on an initial spacelike surface~\cite{COLLQUANNSIN,COLLQUANNSIN1}. This condition requires that there must be no trapping of light at initial spacelike surface from which the collapse starts its evolution. In order to fulfill this condition, trapped surface formation must not take place at the onset of collapse scenario. Thus the regularity of initial spacelike surface requires that $\Theta_+\Theta_-\Big|_{\tau_{\rm i}}\!<0$ or equivalently $R_{\rm i}^2(1+a_{\rm i}^2H^2(a_{\rm i}))<1$.\label{con1} 
	\item For the sake of physical validity of the collapse setting the weak energy condition (WEC) must be satisfied. This condition states that the energy density as measured by any local observer is non-negative. Thus, for energy momentum tensor of ordinary matter the following conditions must be satisfied along any non-spacelike vector field
	\be\label{WECrhop} \rho\geq0,~~~~\rho+p\geq0.
	\ee
	The first inequality requires that $\rho_{\rm i}>0$ while the second one requires $w\geq-1$.\label{con2}
	\item As the collapse proceeds the energy density increases due to contraction of the cloud, hence, at the bounce time we must have $\rho(a_{\rm b})>\rho(a_{\rm i})$. This condition implies $\beta<0$.\label{con3}
	\item As $\lambda(a)$ parameter is a measure of tendency of the geometry (matter fields) to couple with the matter fields (geometry), i.e., mutual interaction between matter and geometry, we expect that this parameter grows at later stages of the collapse so that at the bounce $\lambda(a_{\rm b})>\lambda(a_{\rm i})$.\label{con4}
\end{enumerate}
The solution obtained in Eq.~(\ref{collrate}) exhibits a non-singular collapse process for which the contracting regime {comes to a halt} at a finite amount of comoving time turning then into an expanding regime at the bounce time. In order that this process be physically reliable we require that the model parameters fulfill conditions (\ref{con1})-(\ref{con4}). Figure (\ref{FIG1}) shows the allowed values of the pair $(\beta,w)$ subject to these conditions. We observe that depending on the initial value of energy density it is possible to have different collapse settings for both negative and positive pressures. Figure (\ref{FIG2}) shows the evolution of collapse rate for the allowed values of $(\beta,w)$ parameters as determined in Fig.~(\ref{FIG1}). We observe that the collapse scenario begins its evolution with a negative rate and after passing a finite amount of comoving time the contracting regime (family of red curves) turn into an expanding regime (family of blue curves) at the bounce event where $H(a_{\rm b})=0$. We note that the condition on reality of the square of collapse rate requires that $H^2(a)\geq0$ throughout the collapse process and also at the post-bounce regime.

\par
\section{Exact solution for $\beta=-1$}
In order to realize the time evolution of the collapsing object we can combine Eqs. (\ref{RasFES}) and (\ref{RasFES1}) to find the following differential equation in terms of scale factor
\bea\label{diffeqsf} 
\ddot{a}=\f{a}{2\beta a_{\rm b}^2a_{\rm i}^\beta}\Big[2\beta a_{\rm i}^\beta-\kappa\rho_{\rm i}a_{\rm b}^2(1+w)\left((\beta+2)a^\beta-2a_{\rm b}^\beta\right)\Big],\nn
\eea
where use has been made of expressions (\ref{collrate}) and (\ref{lambdasolfinal}) for the square of collapse velocity, $\dot{a}^2$, and $\lambda(a)$ parameter. We note that this equation can be also derived by taking the time derivative of Eq. (\ref{collrate}) along with substituting for the square of collapse velocity. The above differential equation can be solved using standard numerical techniques, however, exact solutions for the scale factor can be also found for $\beta\in\mathbb{Z}^{-}$. We therefore obtain the following solution for $\beta=-1$
\bea\label{exsol}
a(\tau)={\sf A}(\tau)+{\sf B}\sinh\left[{\sf C}(\tau-\tau_{\rm i})\right],\nn
\eea
where 
\bea\label{AB}
&&{\sf A}(\tau):=\f{1}{2a_{\rm i}a_{\rm b}\kappa\rho_{\rm i}(1+w)-2}\Bigg[a_{\rm i}a_{\rm i}^2\kappa\rho_{\rm i}(1+w)\nn&+&a_{\rm i}\left(\kappa\rho_{\rm i}a_{\rm b}(1+w)(2a_{\rm i}-a_{\rm b})-2\right)\exp\left[{\sf C}(\tau-\tau_{\rm i})\right]\Bigg],\nn
&&{\sf B}:=\f{1}{2-2a_{\rm i}a_{\rm b}\kappa\rho_{\rm i}(1+w)}\Bigg[\kappa\rho_{\rm i}a_{\rm i}a_{\rm b}(1+w)(2a_{\rm i}-a_{\rm b})\nn&+&2a_{\rm b}^2\dot{a}_{\rm i}{\sf C}-2a_{\rm i}\Bigg],\nn
&&{\sf C}:=\f{\sqrt{1-a_{\rm i}a_{\rm b}\kappa\rho_{\rm i}(1+w)}}{a_{\rm b}},
\eea
and the integration constants have been found according to initial conditions $a(\tau_{\rm i})=a_{\rm i}$ and $\dot{a}(\tau_{\rm i})=\dot{a}_{\rm i}$. We note that the latter condition can be set utilizing Eq. (\ref{collrate}) for the initial value of the collapse rate, as $\dot{a}_{\rm i}=a_{\rm i}H(a_{\rm i})$.

\begin{figure}
	\includegraphics[scale=0.23]{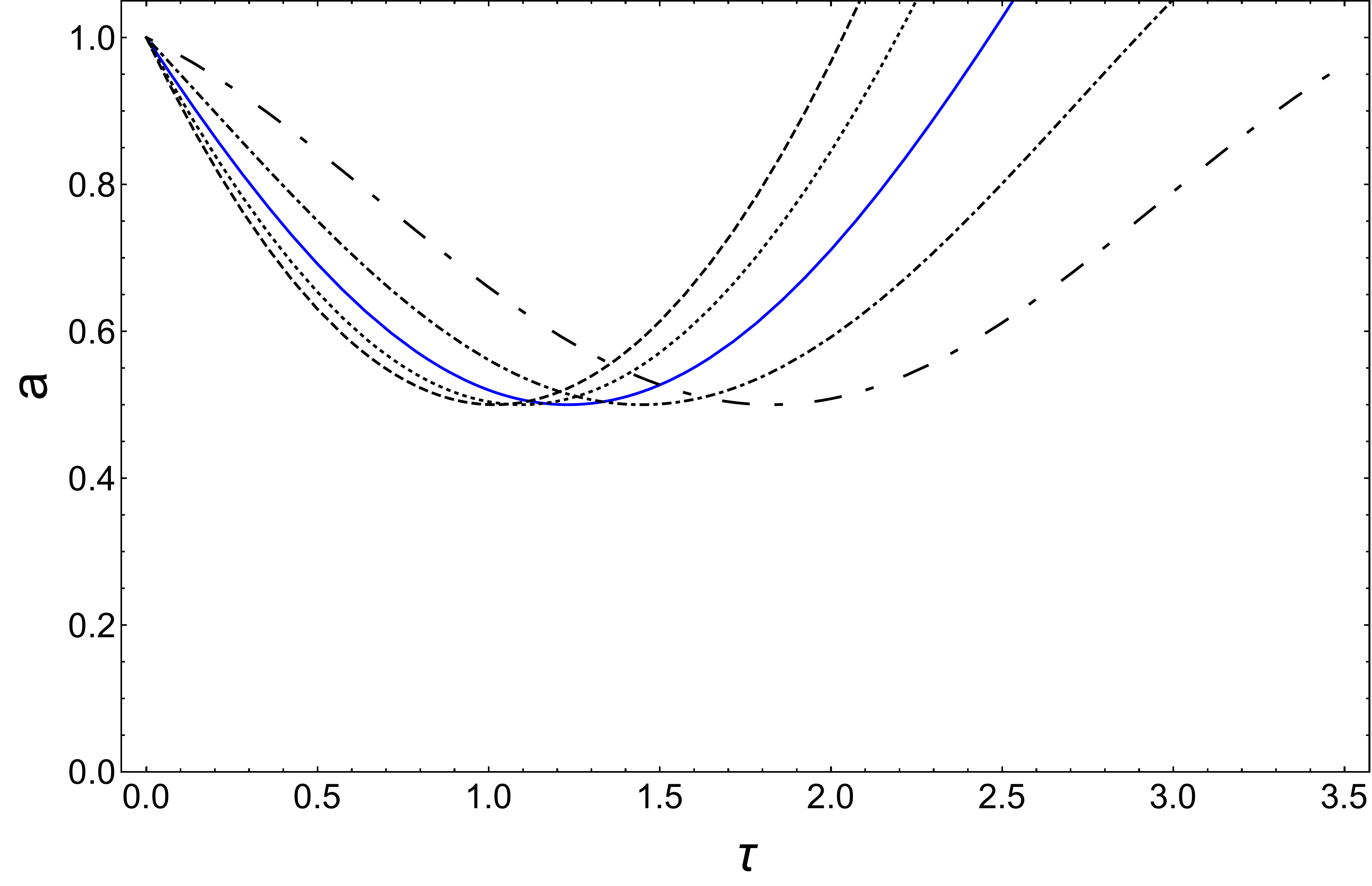}
    \includegraphics[scale=0.23]{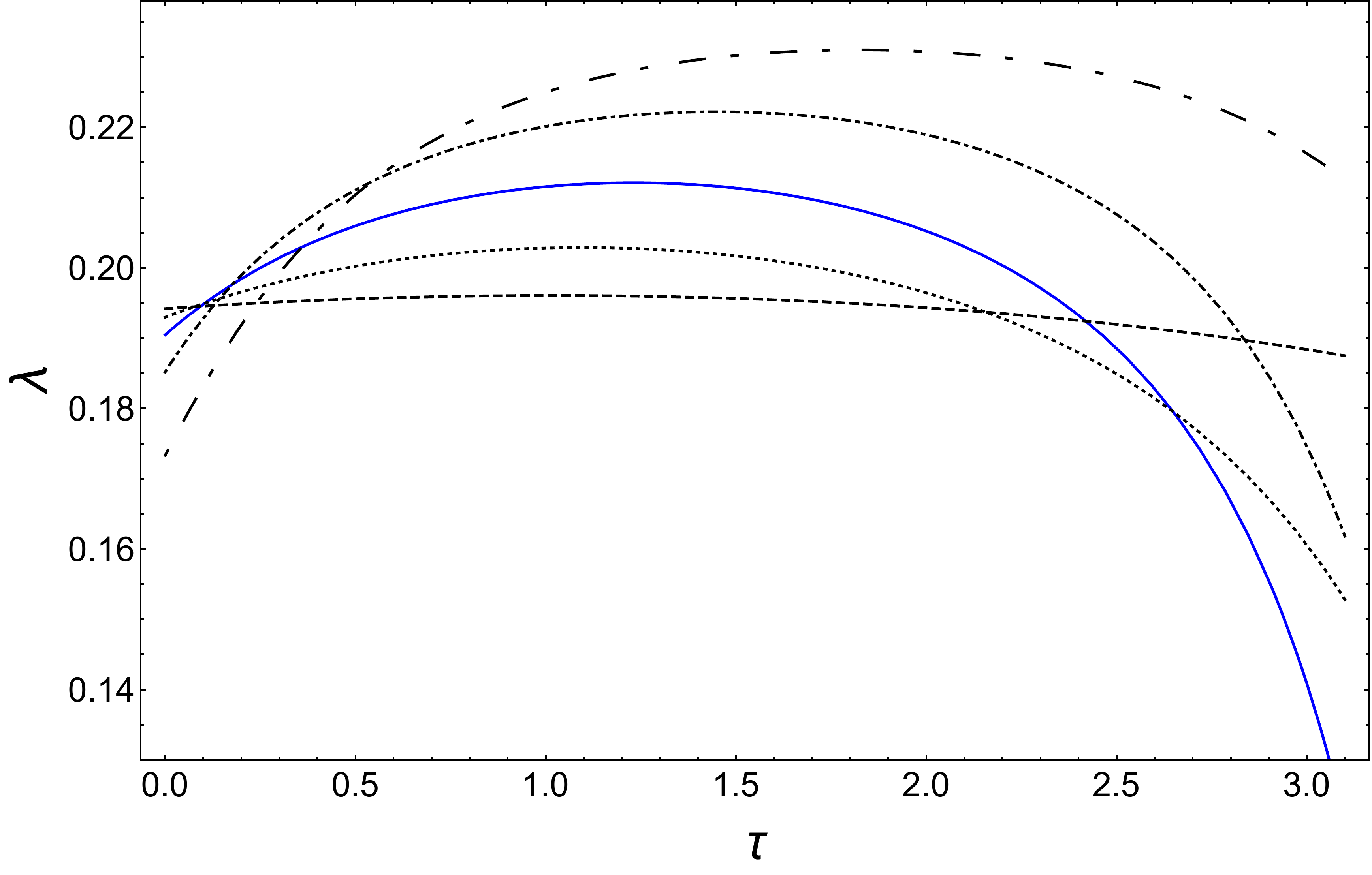}
	\caption{Upper panel: Behavior of scale factor against comoving time for $\beta=-1$, $w=-0.18$ (dashed curve), $w=-0.10$ (dotted curve), $w=0$ (solid curve), $w=0.10$ (dotdashed curve) and $w=0.18$ (long-dotdashed curve) and for $a_{\rm i}=2a_{\rm b}=1$ and $\kappa=1$. Middle panel: Behavior of coupling parameter $\lambda$ against comoving time for the same values of model parameters as of the upper panel.}\label{FIG3}
\end{figure}

In the upper panel of Fig. (\ref{FIG3}) we have sketched the time evolution of the scale factor for $\beta=-1$ and  allowed values of EoS parameter according to Fig. (\ref{FIG1}). We see that the presence of negative pressure ($w<0$) acts against the pull of gravity and prompts the bounce to occur sooner than the case with positive pressure ($w>0$). Though positive pressure tends to delay the occurrence of the bounce, the effects of a dynamic matter-curvature coupling (encoded in $\lambda(\tau)$ parameter) will be dominant near the bounce to finally remove the spacetime singularity. For the case of a pressure-less fluid (solid blue curve) we also see that mutual matter-curvature interaction could prevent the spacetime singularity that forms in OSD collapse model. In the middle panel we have plotted the time behavior of coupling parameter for the same values of model parameters as of the upper panel, where as we expected, the mutual interaction between matter and geometry starts from a finite value reaching a maximum at the bounce time and then monotonically decreases. We note that for $w<0$ the maximum value of coupling parameter is lesser than that of $w>0$ case or equivalently 
\be\label{lambs}
\lambda(a)\big|_{a=a_{\rm b}}^{w<0}<\lambda(a)\big|_{a=a_{\rm b}}^{w=0}<\lambda(a)\big|_{a=a_{\rm b}}^{w>0}.
\ee
In other words, the presence of negative pressure assists the effects of matter-curvature interaction to precede the occurrence of the bounce while positive pressure opposes such effects and therefore, more time is needed in order that the effects of this interaction dominates the gravitational attraction to finally prevent the singularity formation.
\par
\begin{figure}
	\includegraphics[scale=0.22]{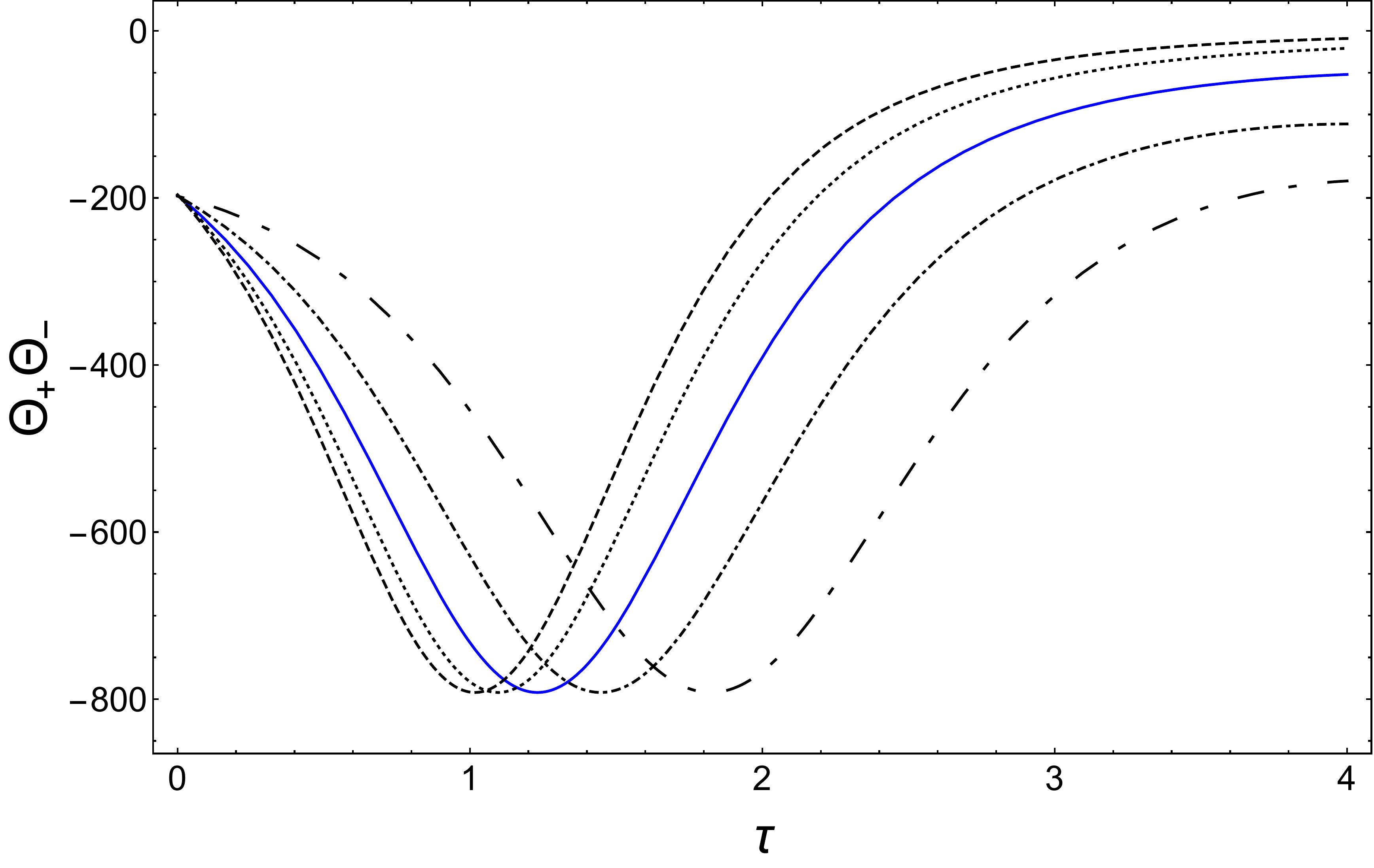}
	\caption{Plot of $\Theta_+\Theta_-$ versus comoving time for $R_{\rm i}=0.1$ and the same model parameters as of Fig (\ref{FIG3}).}\label{FIG4}
\end{figure}
\begin{figure}
	\includegraphics[scale=0.23]{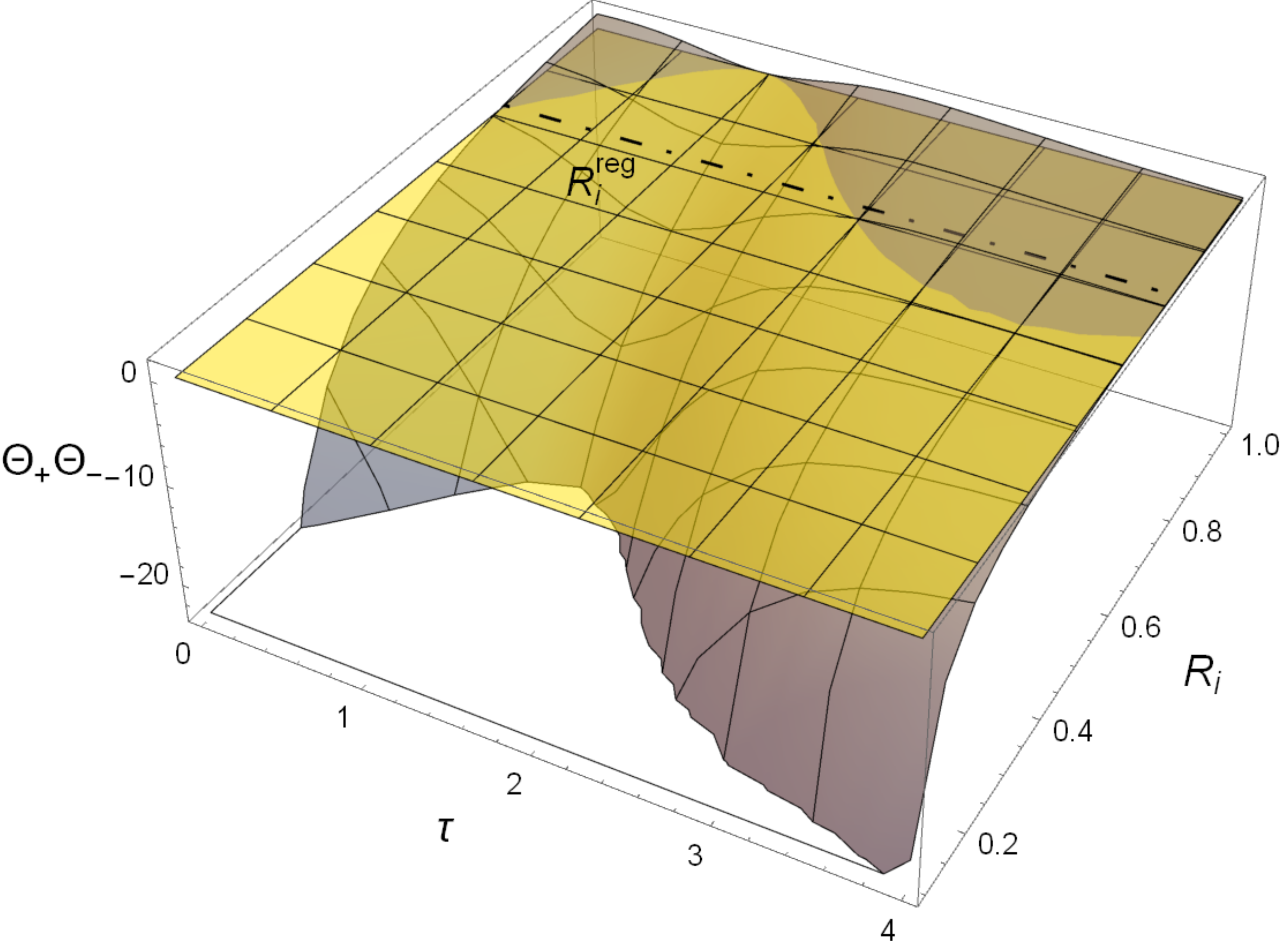}
	\includegraphics[scale=0.23]{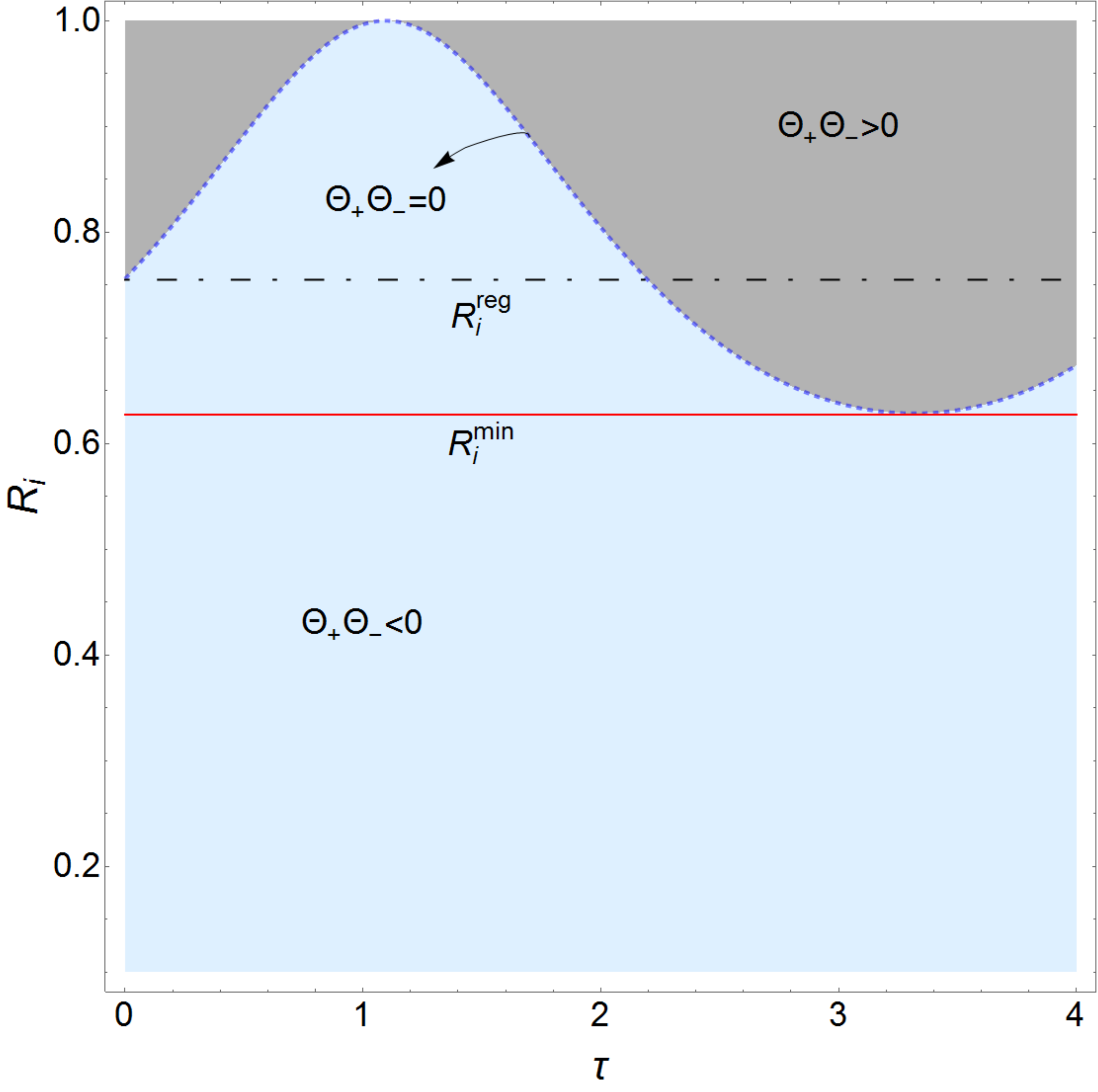}
	\caption{Upper panel: Plot of $\Theta_+\Theta_-$ versus ($\tau,R_{\rm i}$) for $(\beta,w)=(-1,-0.1)$ and the same model parameters as of Fig (\ref{FIG3}). Lower panel: Contour plot of the upper panel. The gray and light blue regions represent the trapped and untrapped regions, respectively.}\label{FIG5}
\end{figure}

Finally, in order to find whether the bounce is visible or not we need to study the dynamics of the apparent horizon. To this aim we consider conditions given in Eq. (\ref{THtruntr}) and evaluate $\Theta_+\Theta_-$, which is given by
\be\label{THTHPM}
\Theta_+\Theta_-=\f{2}{R_{\rm i}^2a^2}\left[R_{\rm i}^2(1+\dot{a}^2)-1\right].
\ee
The behavior of the above expression can be examined for different values of initial radii of the collapsing body. Figure (\ref{FIG4}) shows variation of $\Theta_+\Theta_-$ parameter against comoving time for a specific value of $R_{\rm i}$. We observe that $\Theta_+\Theta_-\Big|_{\tau=\tau_{\rm i}}\!\!\!\!\!\!<0$, and stays negative throughout the dynamical evolution of the collapse scenario. Hence for this specific value of initial radius the apparent horizon formation is avoided. However, one may seek other untrapped bouncing scenarios for different initial radii. In order to get a better understanding of the situation we proceed to evaluate expression (\ref{THTHPM}) for different values of initial radius. The upper panel in Fig. (\ref{FIG5}) shows the evolution of $\Theta_+\Theta_-$ in terms of comoving time and initial radius. We firstly note that this quantity is negative for some values of initial radius of the collapsing object i.e., for $R_{\rm i}<R_{\rm i}^{\rm reg}$ where $R_{\rm i}^{\rm reg}$ (black dot-dashed line) is the minimum allowed value for initial radius for which the regularity of the initial data is preserved, see condition (\ref{con1}). Then, for a suitable choice of initial radius, this quantity can remain negative throughout the contracting and expanding phases and thus, the bounce event is not trapped by the apparent horizon. However, there exist other values of initial radius for which $\Theta_+\Theta_-$ becomes zero and then gets positive values. This situation is better illustrated in the lower panel where we have provided a contour plot representation of the upper panel. The dashed blue curve shows the intersection of the surface $\Theta_+\Theta_-(\tau,R_{\rm i})$ with $\Theta_+\Theta_-(\tau_{\rm ah},R_{\rm i})=0$ plane so that each point on this curve represents the time of apparent horizon formation for its corresponding initial collapse radius. We therefore see that for $R_{\rm i}^{\rm min}<R_{\rm i}<R_{\rm i}^{\rm reg}$ the quantity $\Theta_+\Theta_-$ changes its sign from negative to positive values and thus the apparent horizon will form at a certain time to cover the bounce. However, there exists the minimum radius $R_{\rm i}^{\rm min}$ (red line) so that for $R_{\rm i}<R_{\rm i}^{\rm min}$ the apparent horizon will never meet the boundary of the collapsing object and hence the bounce will not be dressed by the apparent horizon. 

\par
\section{A Comparison to Quantum Corrected Models}\label{QUGRT}
As we mentioned earlier, in the framework of classical GR, singularity theorems predict that under physically reasonable circumstances, such as energy and causality conditions, the occurrence of a spacetime singularity as the collapse final state is inevitable~\cite{HAWPENST,HAWPENST1,Clarke-Tipler,Senovilla1997}. {At classical level, some attempts have been made so far in order to remove the spacetime singularities at the end of a collapse process. On the other hand, a huge amount of work has been devoted to quantum modifications of GR in order to cure the occurrence of classical singularity~\cite{Oriti2009}, see also \cite{MALA2017} for recent reviews.} However, we do not have yet a consistent quantum theory of gravity and attempts made in this way have been focused on investigating some effective theories describing quantum gravity at a semiclassical level by introducing a suitable modification of GR in the strong field regime. Work along this line has been performed within cosmological~\cite{QGCOSMOL}, as well as astrophysical scenarios~\cite{ALTGRNONSIN}, where in the former, it is shown that the initial singularity of the Universe can be resolved and replaced by a non-singular bounce. In the latter studies, loop quantum gravity (LQG) has been utilized in order to resolve the spacetime singularity that would arise as the final state of a classical collapse scenario. For example, in~\cite{Tavakoli2014} LQG corrections to the collapse of a mass-less scalar field have been investigated and it is shown that such corrections can provide a setting to avoid the singular state of classical collapse, see also~\cite{LQGCOLL} for further studies on this issue.
\par
In this section we compare the solution obtained in Eq.~(\ref{collrate}) for the dust case with quantum gravity induced corrections to a classical collapse setting which is introduced in~\cite{Tavakoli2014,Bambi2013}. Using then Eq. (\ref{energa}) for energy density, Eq. (\ref{collrate}) can be rewritten as
\be\label{H2R}
H^2(\rho)=\f{1}{a_{\rm b}^2}+\f{\rho_{\rm i}\kappa}{\beta}\left(\f{a_{\rm b}}{a_{\rm i}}\right)^\beta-\f{\kappa\rho}{\beta}-\f{1}{a_{\rm i}^2}\left(\f{\rho_{\rm i}}{\rho}\right)^{\f{2}{\beta}}.
\ee
In the framework of an effective theory of gravity, quantum corrections lead to the modified Friedmann equation 
\be\label{rhocorr}
H^2(\rho)=\f{\rho}{3}\left(1-\f{\rho}{\rho_{\rm cr}}\right),
\ee
where $\rho_{\rm cr}\approx0.41\rho_{{\rm Pl}}$ and $\rho_{{\rm Pl}}$ is the Planck energy density. Even if we merely consider a classical picture for the collapse scenario in GRT, i.e., $a_{\rm b}\gg\ell_{\rm Pl}$ and $\rho_{\rm i}<\rho\ll\rho_{{\rm Pl}}$ then the collapse rate in Eq. (\ref{H2R}) vanishes at $a=a_{\rm b}$ with $(\rho_{\rm i},\rho_{\rm b})$ given by
\be\label{classicrho}
{\rm for}~~~\beta=-1,~~~~\rho_{\rm i}=\f{1}{\kappa a_{\rm i}a_{\rm b}},~~~~~\rho_{\rm b}=\f{1}{\kappa a_{\rm b}^2}.
\ee
Then, a contracting regime will be changed by an expanding one and as the result, spacetime singularity will be replaced by a non-singular bounce. From another side, if the collapse scenario continues to a Planck scale regime so that ($\rho$,$a_{\rm b}$) are comparable to Planck energy density and Planck length, then, by comparing Eq.~(\ref{rhocorr}) with Eq.~(\ref{H2R}) we find that the modified Friedmann equation in GRT mimics its quantum counterpart in the following way
\be\label{EQSCOMP}
H^2(\rho)=\f{\rho}{a_{\rm b}^2\rho_{\rm cr}}\left(1-\f{\rho}{\rho_{\rm cr}}\right),
\ee
where we have set the model parameters as
\be\label{modpars}
\beta=-1,~~~\kappa=\f{1}{a_{\rm b}^2\rho_{\rm cr}},~~~\rho_{{\rm i}}=\f{a_{\rm b}}{a_{\rm i}}\rho_{\rm cr},
\ee
and the constant $1/3$ has been absorbed into $\kappa$ parameter. We therefore observe that a dynamic matter-curvature interaction, as described by GRT, has the ability to play the role of corrections due to the quantum gravity effects within the energy density.
\section{concluding remarks}\label{ConC}
It is generally expected that quantum effects of gravity that may appear in the late stages of a gravitational collapse scenario could resolve the spacetime singularities in classical collapse setting where all known laws of physics breakdown. However, attempts toward singularity avoidance at classical level have shown that it is still possible to prevent the singularity formation within the framework modified gravity theories. In this respect additional contributions to GR action can assist removing the singularity without resorting to quantum effects. Our aim in the present study was to examine the possibility of singularity removal for a classical collapse setting in the context of GRT. We therefore observed that the collapse process in GRT follows a quite different path as compared to the collapse scenario {within the framework of RG}~\cite{PLB}. The exact solutions in the herein model represent a dynamical evolution for the collapse process that does not terminate in a spacetime singularity. Indeed, the contracting phase which begins from regular initial data turns into an expanding one at {the} bounce time $\tau=\tau_{\rm b}$, hence the formation of spacetime singularity is avoided. {We further observed that as the collapse advances, $\lambda(a)$ parameter grows until reaching a maximum value at the bounce time, where the collapse halts $(\dot{a}(\tau_{\rm b})=0)$ at a minimum value of the scale factor. As this parameter is a measure of mutual running interaction between matter and geometry, one may then intuitively imagine that at late stages of the collapse scenario, gravitational attraction succumbs to the effects of such an interaction leading finally to singularity removal.} It is {also} worth mentioning that, as $a_{\rm b}\gg\ell_{\rm Pl}$, the fate of a collapse setting with singular endstate may be altered long before quantum gravity effects become important. {We therefore conclude that}, at classical level, the effects of {a varying } nonminial coupling between matter and geometry (encoded in {the} dynamic Rastall parameter) could act as an agent for singularity {avoidance} in gravitational collapse of a homogeneous perfect fluid. 
\acknowledgments{The authors would like to appreciate the anonymous referee for providing useful and constructive comments that helped us to improve the original version of our manuscript.} 

\end{document}